\documentclass[a4paper, 10pt, conference]{ieeeconf}      
\usepackage{amsmath, amssymb}
\usepackage{graphicx}
\usepackage{subfigure}
\usepackage[noadjust]{cite}
\usepackage{color}
\usepackage{amsmath, amssymb}
\usepackage{graphicx}
\usepackage{subfigure}
\usepackage[noadjust]{cite}
\usepackage{color}
\usepackage[noadjust]{cite}
\usepackage{color}
\usepackage{wrapfig}

\usepackage{algorithmic}
\usepackage{algorithm}

\usepackage{hyperref}
\usepackage{mathrsfs}
\usepackage{cite}
\IEEEoverridecommandlockouts                              
\title{\huge \bf
Safe Reinforcement Learning for Emergency Load Shedding of Power Systems
}

\author{Thanh Long Vu,~\IEEEmembership{Member,~IEEE,} Sayak Mukherjee,~\IEEEmembership{Member,~IEEE,} Tim Yin, \\ Renke Huang,~\IEEEmembership{Member,~IEEE,} Jie Tan, 
and~Qiuhua Huang,~\IEEEmembership{Member,~IEEE}
\thanks{T. L. Vu, T. Yin, S. Mukherjee, R. Huang, Q. Huang are with the Pacific Northwest National Laboratory, 902 Battelle Blvd, Richland, WA 99354 USA, e-mail: \texttt{thanhlong.vu@pnnl.gov}. J. Tan is with Google Brain, Google Inc, Mountain View, CA, 94043 USA (e-mail: \texttt{jietan@google.com}).
}}

\begin{document}

\maketitle
\thispagestyle{empty}
\pagestyle{empty}

\maketitle
\begin{abstract}
    The paradigm shift in the electric power grid necessitates a revisit of existing control methods to ensure the grid's security and resilience. In particular, the increased uncertainties and rapidly changing operational conditions in power systems have revealed  outstanding issues in terms of either speed, adaptiveness, or scalability of the existing control methods for power systems. On the other hand, the availability of massive real-time data can provide a clearer picture of what is happening in the grid. Recently, deep reinforcement learning (RL) has been regarded and adopted as a promising approach leveraging massive data for fast and adaptive grid control. However, like most existing machine learning (ML)-based control techniques, RL control usually cannot guarantee the safety of the  systems under control. In this paper, we introduce a novel method for safe RL-based load shedding of power systems that can enhance the safe voltage recovery of the electric power grid after experiencing faults. Numerical simulations on the $39$-bus IEEE benchmark is performed to demonstrate the effectiveness of the proposed safe RL emergency control, as well as its adaptive capability to faults not seen in the training. 
\end{abstract}

\section{Introduction}
The  electric power grid, the largest man-made system, has been experiencing a transformation to an even more complicated
system with an increased number of distributed energy sources and more active and less predictable load endpoints. At the same time, intermittent renewable generation introduces high uncertainty into system operation and may compromise power system stability and security. The existing control operations and modeling approaches, which are largely developed several decades ago for the much more predictable operation of a vertically integrated utility with no fluctuating generation, need to be reassessed and adopted to more stressed and rapidly changing operating conditions \cite{camacho2011control, Chen2013139}. \textcolor{black}{In particular, operating reserves \cite{Wood1996} face limitations in the current grid paradigm. 
Hence, emergency control, i.e., quick actions to recover the stability of a power grid  under critical contingency, is more frequently required.}

Currently, emergency control of power grids is largely based on remedial actions, special protection schemes (SPS), and load shedding \cite{Vittal2003}, which aim to quickly rebalance power and hopefully stabilize the system. These emergency control measures historically make the electrical power grid reasonably stable to disturbances. However, with the increased uncertainties and rapidly changing operational conditions in power systems, the existing emergency control methods have exposed outstanding issues in terms of either speed, adaptiveness, or scalability. Load shedding is well known, among the measures of emergency voltage control, as an effective countermeasure against voltage instability \cite{Taylor92}. It has been widely adopted in the industry, mostly in the form of rule-based undervoltage load shedding (UVLS). The UVLS relays are usually employed to shed load demands at substations in a step-wise manner if the monitored bus voltages fall below the predefined voltage thresholds. ULVS relays have a fast response, but do not have communication or coordination among substations, leading to unnecessary load shedding \cite{Bai11} at affected substations. 

Recently, data-driven approaches have been getting investigated to tackle the voltage instability phenomenon in ways less dependent on the model and more adaptive to contingencies. \cite{Genc2010_DT} proposed a decision tree based approach for preventive and corrective control actions. For load shedding against fault-induced delayed
voltage recovery (FIDVR) events, \cite{Qiao2020hierarchical_LS} developed a hierarchical extreme-learning machine based algorithm. Reinforcement learning (RL) based approaches are also applied to voltage control problem in works such as \cite{huang2019loadshedding_DRL, zhang2018loadshedding_DRL, huang2020accelerated}. In this approach, the machine agent interacts with the system/ environment, observes the resulting system state and the corresponding reward, and updates the control actions in a way to maximize the reward which is defined to encourage the voltage stability constraint \cite{huang2019loadshedding_DRL}. 

Over the last decade, a varieties of RL algorithms have been developed, which mainly use the Markov decision process (MDP) based framework. In our previous work, we have designed a deep Q learning based RL control \cite{huang2019loadshedding_DRL} for emergency load shedding in response to voltage stability issues of the electric power grid. Subsequently, we developed an accelerated Augmented Random Search (ARS) algorithm in \cite{huang2020accelerated} to quickly train the neural network-based control policy in a parallel computing mechanism. This approach, sketched in Section \ref{sec.ARS}, is implemented on a high-performance computing platform in a novel nested parallel architecture using the Ray framework. This architecture allowed parallelizing power grid dynamic simulations in the ARS training and  helped exploring the parameter space of the control policy efficiently and adapting ARS to multiple tasks.

However, one of the limitations of most existing reinforcement learning methods is that they usually cannot guarantee the safety of the system. In this paper, we develop a safe ARS algorithm that can enhance the safety of the power grid during training, while inheriting all the advantages of the standard ARS algorithm. In this safe ARS method, as described in Section \ref{sec.safeARS}, the safety requirement of the power grid in load shedding is defined by a constraint on a time-dependent safety function. Accordingly, goal of the ARS agent is to learn the optimal control policy that maximizes the expected reward, and at the same time, satisfies the safety requirement defined by the constraint on the safety function. To solve this constrained optimization, we formulate a Lagragian function involving both normal reward function and the safety function with some multiplier. Subsequently, the safe ARS agent learns an optimal emergency control policy to maximize the Lagrangian function over the space of policies and minimize this function over the value space of the safety multipliers. We demonstrate the effectiveness of the resulting safe ARS-based emergency load shedding on the 39-bus IEEE testcase in Section \ref{sec.results}. Furthermore, we show that in comparison to the standard ARS-based load shedding \cite{huang2020accelerated}, the safe ARS-based load shedding can be more adaptive to contingencies not seen in the training, indicating its promise in ensuring the safe operation of actual power systems in deployment.

\section{Machine Learning-based Emergency Load Shedding}
\label{sec.ARS}

\begin{figure}[t!]
    \centering
    \includegraphics[width = 1.0\linewidth, height = 5.5 cm]{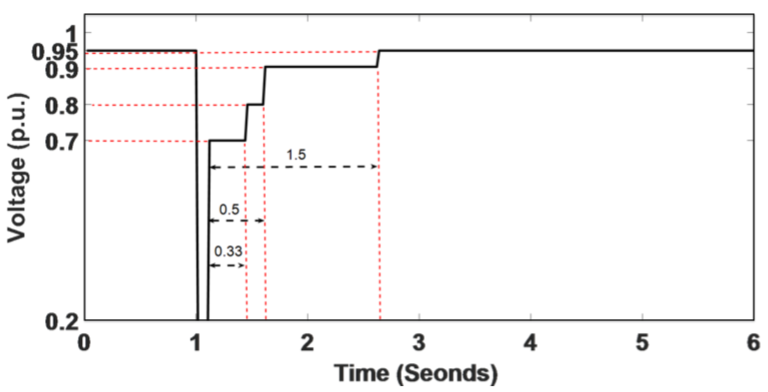}
    \caption{Transient voltage recovery criterion for transmission system}
    \label{fig.requirement}
    \end{figure}
In the load shedding problem, the objective is to recover the voltages of the electric power grid after the faults so that the post-fault voltages  will recover according to the standard [13] showed in Figure \ref{fig.requirement}. 
In particular, the standard requires that, after fault clearance, voltages should return to at least $0.8, 0.9$, and $0.95$ p.u. within $0.33s, 0.5s$, and $1.5s$. In our previous work, we have designed a deep Q learning based RL control \cite{huang2019loadshedding_DRL} for emergency load shedding to address this problem. Subsequently, we developed an accelerated Augmented Random Search (ARS) algorithm in \cite{huang2020accelerated} to quickly train the neural network-based control policy in a parallel computing mechanism. The common root of these two works is the reinforcement learning (RL), which was extensively developed last some decades using the Markov decision process (MDP) based framework. 

In the general RL framework, an RL agent interacts with the system/ environment, observes the resulting system state and the corresponding reward, and updates the control actions in a way to maximize the reward. Mathematically, we define a policy search problem in a (partially observable) MDP defined by a tuple $(S,A, \mathcal{P},r,\gamma)$ \cite{SuttonBarto}. The state space $S$ and action space $ A$ could be continuous or discrete. In this paper, both of them are continuous. The environment transition function $\mathcal{P}: S \times A \to S$  is the probability density of the next state $s_{t+1} \in S$ given the current state $s_t \in S$ and action $a_t \in A$. At each interaction step, the environment returns a real reward $r: S \times A \to R$. $\gamma \in (0,1)$ is the discount factor. The standard RL objective is to maximize the expected sum of discounted rewards.
Many RL algorithms have been developed in the literature such as 
value iteration based algorithms like deep Q learning 
policy gradient based algorithms like deep deterministic policy gradient (DDPG) 
optimization influenced algorithms such as trust region policy optimization (TRPO) 
actor-critic methods such as soft actor-critic (SAC), etc.


In the ARS algorithm, that we used in \cite{huang2020accelerated}, the ARS agent performs parameter-space exploration and estimates the gradient of the expected reward using sampled rollouts to update the control policy. In particular, for the above load shedding problem, the ARS agent's objective is to maximize the expected reward, where
the reward $r_t$ at time $t$ was defined as follows:
\begin{align}
\label{eq.reward}
r(t) = 
\begin{cases}
&-1000  {\;\;\; \emph{if}\;\;\;} V_i(t)<0.95,\;\;\; T_{pf} +4<t \\
& c_1 \sum_i \Delta V_i(t) - c_2 \sum_j \Delta P_j (p.u.) - c_3 u_{ivld}, \\&{\;\;\; \emph{\mbox{otherwise}},}
\end{cases}
\end{align}
where,
\footnotesize
\begin{align}
\Delta V_i(t) =
\begin{cases}
&\min \{V_i(t) - 0.7, 0\},  \emph{\;if\;} t \in (T_{pf}, T_{pf} +0.33) \\
&\min \{V_i(t) - 0.8, 0\}, \emph{\;if\;} t \in (T_{pf} +0.33, T_{pf} +0.5 ) \\
&\min \{V_i(t) - 0.9, 0\} , \emph{\;if\;}  t \in (T_{pf} +0.5, T_{pf} +1.5) \\
&\min \{V_i(t) - 0.95, 0\} , \emph{\;if\;} t \in t > T_{pf} +1.5.
\end{cases}\nonumber
\end{align}
\normalsize

In the reward function \eqref{eq.reward}, $T_{pf}$ is the time instant of fault clearance;
$V_i(t)$ is the voltage magnitude for bus $i$ in the power
grid;  $\Delta P_j (t)$ is the
load shedding amount in p.u. at time step $t$ for load bus $j$; 
invalid action penalty $u_{ivld}$ if the DRL agent still provides
load shedding action when the load at a specific bus has already been shed to zero at the previous time step when
the system is within normal operation. $c_1 , c_2$, and $c_3$ are the weight factors for the above three parts.

 Furthermore, to scale up the ARS algorithm and reduce the training time, we accelerated it by leveraging its inherent parallelism and implementing it on a high-performance computing platform in a novel nested parallel architecture using the \textit{Ray} framework. This architecture allowed parallelizing power grid dynamic simulations in the ARS training and helped exploring the parameter space of the control policy efficiently and adapting ARS to multiple tasks \cite{huang2020accelerated}.

\section{Safe ARS-based Emergency Load Shedding}
\label{sec.safeARS}

The safety requirement of the power grid can be defined by a constraint on a time-dependent safety function $f(s_t,a_t,t)$, i.e., the system is safe if $f(s_t,a_t,t) \ge 0$. For the voltage safety requirement in Figure \ref{fig.requirement}, the following time-dependent safety function can be used:

\begin{align}
\label{safeVol}
   f(s_t,a_t,t)= \begin{cases}
   & 0.4^2 -\max_i (V_i(t)-1.1)^2 \\& {\;\;\; \emph{if}\;\;\;} T_{pf}<t<T_{pf}+0.33, \\
   & 0.35^2 -\max_i (V_i(t)-1.15)^2 \\& {\;\;\; \emph{if}\;\;\;} T_{pf}+0.33<t<T_{pf}+0.5, \\
   &0.3^2 -\max_i (V_i(t)-1.2)^2  \\&{\;\;\; \emph{if}\;\;\;} T_{pf}+0.5<t<T_{pf}+1.5, \\
   &0.275^2 -\max_i (V_i(t)-1.225)^2  \\&{\;\;\; \emph{if}\;\;\;} t>T_{pf}+1.5. 
   \end{cases}
\end{align}
where $V_i(t)$ is the bus voltage magnitude for bus $i$ in the power
grid. The constraint $f(s_t,a_t,t) \ge 0$ will ensure the voltage will recover as in the criterion in Figure \ref{fig.requirement}, while always less than $1.5 p.u.$
The goal of a safe ARS agent is to learn the optimal control policy that maximizes the expected reward, and at the same time, satisfies the safety requirement defined by $f(s_t,a_t,t) \ge 0$. A natural approach  that was extensively used in the literature \cite{JMLR:v16:garcia15a} is to formulate this problem as the following constrained optimization  
\begin{align}
    &\max_{a_t \in A} E[ r(s_t,a_t)] \\
    \mbox{s.t.} \;\; & E[f(s_t,a_t,t))] \ge 0.
\end{align}
which can be approximately solved by putting a high penalty for the safety violation \cite{KADOTA2006279}.  

In this paper, we proposed a different approach in which we consider the Lagrangian function  
\begin{align}
    \mathbf{L}(\pi,\lambda) &= E[ r(s_t,a_t)]  + \lambda E[f(s_t,a_t,t)] \\
    \label{Lfunction}& = E[ r(s_t,a_t) + \lambda f(s_t,a_t,t)] 
\end{align}
where $\lambda>0$ is the safety multiplier. 
Since in the original optimization we have the constraint $f(s_t,a_t,t) \ge 0$, the dual function $\mathbf{L}(\pi,\lambda)$ is an upper bound for the original optimization problem. As such, we can use the dual gradient descent method to get the optimum value of both $\mathbf{L}(\pi,\lambda)$ and the safety multiplier $\lambda$, in which we maximize $\mathbf{L}(\pi,\lambda)$ when $\lambda$ is fixed as in the standard ARS algorithm and minimize $\mathbf{L}(\pi,\lambda)$ by using gradient method to find the optimal value of $\lambda$. In other words, the objective of the safe ARS agent is to learn an optimal policy $\pi^*_{\theta}(s_t,a_t)$ from a {min-max} problem on the Lagrangian function
\begin{align}
    (\pi^*, \lambda^*) = \bf{argmax}_{a \in A} \bf{argmin}_{\lambda>0} \mathbf{L}(\pi,\lambda)
\end{align}




For simplicity, we also proposed a more heuristic approach to find the suboptimal value of the safety multiplier $\lambda$, in which we will check the safety conditions in each iteration. If the safety constraint is not violated in the iteration, we reduce the value of $\lambda$ two times. Otherwise, we increase the value of $\lambda$ two times.  Algorithm $1$ presents the steps to compute the optimal control policy and safety multiplier. 
\begin{algorithm}[]
\caption{Safe ARS (sARS) for Load Shedding}
\begin{algorithmic}
\STATE 1. \textbf{Hyperparameters:} Step size $\alpha$, number of policy perturbation directions per iteration $N$, standard deviation of the exploration noise $\nu$, number of top-performing perturbed directions selected for updating weights $b$, number of rollouts per perturbation direction $m$. Decay rate $\epsilon$. 
\STATE 2. \textbf{Initialize:} Policy weights $\theta_0$ with small random numbers; initial safety multiplier in the reward function $\lambda_0,$ the running mean of observation states $\mu_0 = 0 \in R^n$ and the running standard deviation of observation states $\Sigma_0 = I_n \in R^n$, where $n$ is the dimension of observation states, the total iteration number $H$.
\FOR{ iteration $t \leq H$}
\STATE 3. Sample $N$ number of random directions $\delta_{1},\dots,\delta_{N} \in \mathbb{R}^{n_{\theta}}$ with the same
dimension as policy weights $\theta$.
\FOR{each $\delta_{i}, i=1,\dots, N$}
\STATE 4. Add perturbations to policy weights: $\theta_{ti+} = \theta_{t - 1} +
\nu \delta_i$ and $\theta_{ti-} = \theta_{t - 1} - \nu \delta_i$
\STATE 5. Do total $2m$ rollouts (episodes) denoted by $R_{p\in T} (.)$
for different tasks $p$ sampled from task set $T$ corresponding to $m$ different faults with the $\pm$ perturbed policy weights. Calculate safety functions for all $\pm$ perturbations and calculate the average rewards of $m$ rollouts as the rewards for $\pm$ perturbations, i.e., $\bar{R}_{ti +}$ and $\bar{R}_{ti -}$ are 
\begin{align}
    &\bar{R}_{ti+} = \frac{1}{m} R_{p\in T}(\theta_{ti+},\mu_{t - 1},\Sigma_{t - 1}),\\
    &\bar{R}_{ti-} = \frac{1}{m} R_{p\in T}(\theta_{ti-},\mu_{t - 1},\Sigma_{t - 1})
\end{align}
where the reward function $R$ is defined as the combined reward function in \eqref{Lfunction}, i.e., $R(.) = r(s_t,a_t) + \lambda_t f(s_t,a_t,t).$
\STATE 6. During each rollout, states $s_{t,k}$ at time step $k$ are first normalized and then used as the input for inference with policy $\pi_{\theta_t}$ to obtain the action $a_{t,k}$, which is applied to the environment and new states $s_{t,k+1}$ is returned, as shown in (3). The running mean $\mu_t$ and standard deviation $\Sigma_t$ are updated with $s_{t,k+1}$
\begin{align}
    &s_{t,k} = \frac{s_{t,k} - \mu_{t-1}}{\Sigma_{t-1}},\\
    & a_{t,k} = \pi_{\theta_t}(s_{t,k}),\\
    & s_{t,k+1} \gets \mathcal{P}(s_{t,k},a_{t,k}) ,
\end{align}
\ENDFOR
\STATE 7. Sort the directions based on $\max ( \bar{R}_{ti + } , \bar{R}_{ti - } )$, select
top $b$ directions, calculate their standard deviation $\sigma_b$
\STATE 8. Update the policy weight:
\begin{align}
    \theta_{t+1} = \theta_{t} + \frac{\alpha}{b  \sigma_{b}} \sum_{i=1}^{b} (\bar{R}_{ti+} - \bar{R}_{ti-})\delta_{i}
\end{align}
\STATE 9. Step size $\alpha$ and standard deviation of the exploration noise $\nu$ decay with rate $\epsilon$: $\alpha \gets \epsilon \alpha, \nu\gets \epsilon \nu$.
\STATE 10. Update of the safety multiplier: Check the safety violation for all the rollouts. If there is any safety violation: $\lambda_{t+1} \gets 2 \lambda_t$. Otherwise: $\lambda_{t+1} \gets \lambda_t /2.$
\ENDFOR
\RETURN 11.  Return $\theta$,  and {$\lambda$}.
\end{algorithmic}
\end{algorithm}

Similar to the standard ARS algorithm \cite{huang2020accelerated}, the safe ARS learner is an actor at the top to delegate tasks and collect returned information, and controls the update of policy weights $\theta$ and the safety multiplier $\lambda$. The learner communicates with subordinate workers and each of these workers is responsible for one or more perturbations (random search) of the policy weights as in Step 4. In Steps 7, the ARS learner combines the results from each worker calculated in Step 5 (which include the average reward of multiple rollouts and safety violation check), sorts the directions according to the reward, selects the  best-performing directions. Then, in Step 8, the ARS learner updates the policy weights and the safety multiplier $\lambda$ centrally based on the perturbation results from the top performing workers. The workers do not execute environment rollout tasks by themselves. They spawn a number of actors and assign these tasks to these subordinate  actors. Note that each worker needs to collect the rollout results (which include the reward function, safety function, and safety violation check) from multiple tasks inferring with the same perturbed policy, and each actor is only responsible for one environment rollout with the specified task and perturbed policy sent by its up-level worker. For the environment rollouts, power system dynamic simulations are performed in parallel.  

\section{Test Results}
\label{sec.results}
\begin{figure}[t!]
    \centering
    \includegraphics[width = .8\linewidth, height = 5.5 cm]{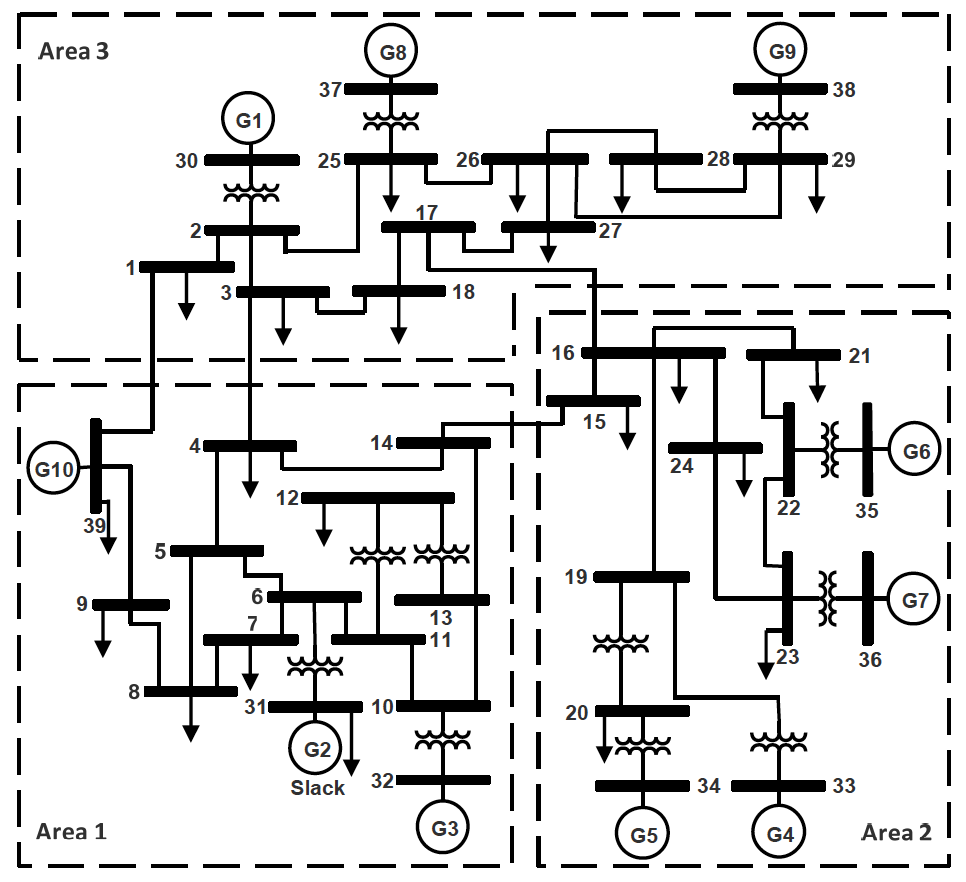}
    \caption{IEEE 39-bus system}
    \label{fig:39bus}
    \end{figure}

We consider the IEEE benchmark $39-$bus, $10-$generator model, showed in Fig. \ref{fig:39bus}. The simulations were performed in the Linux mainframe with $27$ cores. The power system simulator was implemented in the GridPack\footnote{https://www.gridpack.org}, 
and the safe deep RL algorithm was implemented in the python platform. We built a software setup such that the grid simulations in the GridPack and the RL iterations in the python can coordinate.  

We tested the performance of our safe ARS algorithm on the IEEE $39$-bus system to learn a closed-loop control policy for applying the load shedding at a load center including buses $4, 7,$ and $18$ to avoid the FIDVR and meet the voltage recovery requirements shown in Fig. 1. 
Observations included voltage magnitudes at buses $4, 7, 8,$ and $18$ as well as the remaining fractions of loads served by buses $4, 7$ and $18$. The control action for buses $4, 7,$ and $18$ at each action time step was a continues variable from $0$ (no load shedding) to $-0.2$ (shedding $20\%$ of the initial total load at the bus). During the training, the task set T was defined as nine different tasks (fault scenarios). Each task began with a flat start of dynamic simulation. At $1.0$ s of the simulation time a short-circuit fault was applied at bus $4, 15$, or $21$ with a fault duration of $0.0$ s (no fault), $0.15$ s, or $0.28$ s and the fault was self-cleared. 
In \cite{huang2020accelerated}, we observed that the normal ARS can offer robust and effective load shedding to safeguard the system in many contingencies.
In this paper, to test the effectiveness of the safe ARS-based load shedding in comparison with the normal ARS-based load shedding, we consider relatively severe faults that can harm the system. As such, we consider relatively large fault duration. 

\begin{figure}[t]
    \centering
    \includegraphics[width = .9\linewidth, height = 5.5 cm]{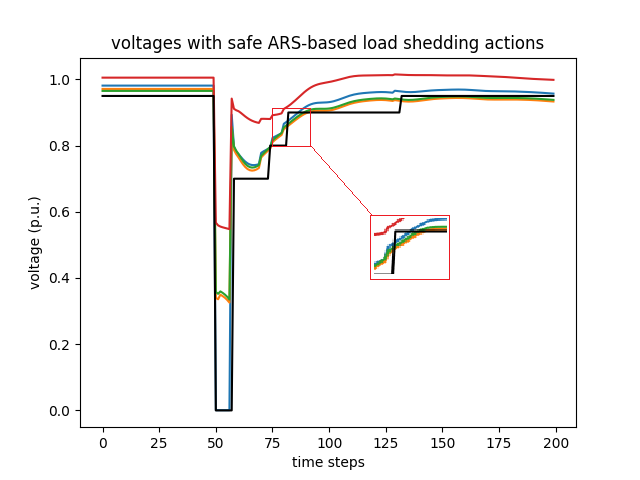}
    \caption{Standard ARS-based emergency load shedding response to the fault that happens at bus 4}
    \label{fig:ARS4}
\end{figure}

\begin{figure}[t]
    \centering
    \includegraphics[width = .9\linewidth, height = 5.5 cm]{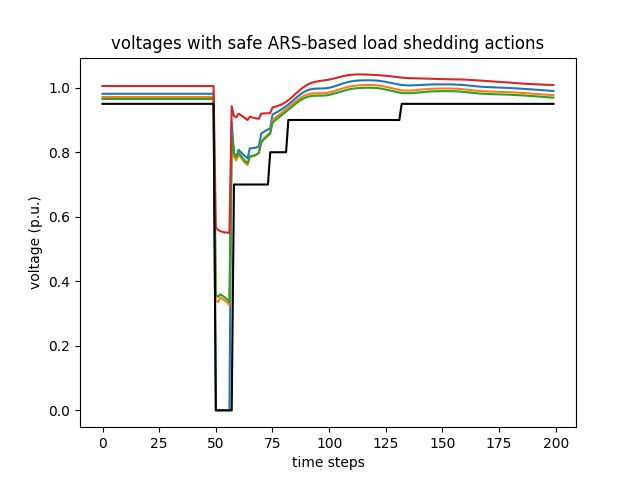}
    \caption{Safe ARS-based emergency load shedding response to the fault that happens at bus 4.}
    \label{fig:safeARS4}
\end{figure}

In the deployment of safe ARS-based emergency control, we consider a short-circuit fault happens at bus $4$ at $1.0$s with a fault duration of $0.15$s and then the fault self-cleared. For comparison with the standard ARS-based emergency control, we also train it with the same set of faults and deploy it to react the same fault at bus $4$. The performances of standard ARS-based load shedding and safe ARS-based load shedding are depicted  in Figs. \ref{fig:ARS4} and \ref{fig:safeARS4}. 
From these figures, we can observe that the safe ARS-based load shedding is better than the standard ARS-based load shedding in meeting the safety requirement described by the transient voltage recovery criterion depicted in Fig. 1. In particular, with the normal ARS-based load shedding, the voltages at buses $7,8,$ and $18$ do not recover to $0.9$ p.u. within $0.5$s after the fault-cleared time, and the voltages at buses $8$ and $18$ do not recover to $0.95$ p.u. within $1.5$s after the fault-cleared time. Yet, the safe-ARS load shedding makes the voltages at all buses $4,7,8,18$ recover well as required. 

\begin{figure}[]
    \centering
    \includegraphics[width = .9\linewidth, height = 5.5 cm]{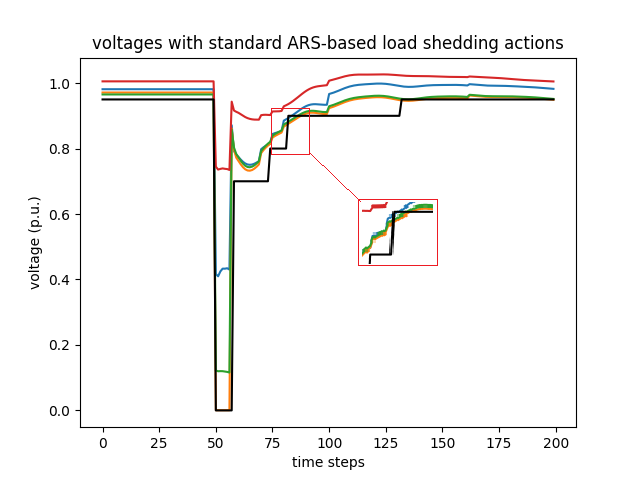}
    \caption{Standard ARS-based emergency load shedding response to the untrained fault that happens at bus 7}
    \label{fig:ARS}
\end{figure}
\begin{figure}[]
    \centering
    \includegraphics[width = .9\linewidth, height = 5.5 cm]{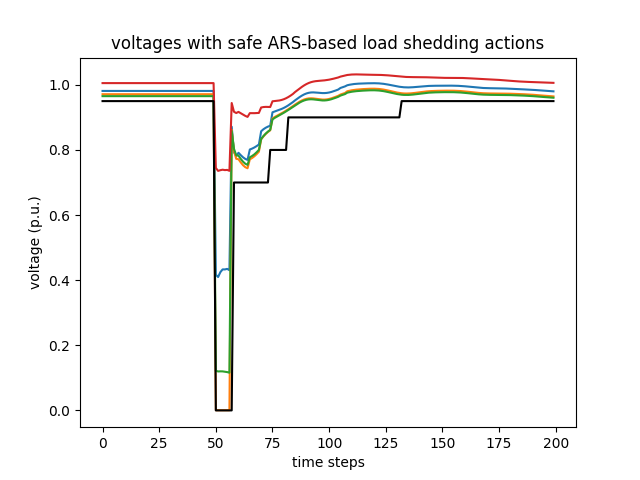}
    \caption{Safe ARS-based emergency load shedding response to the untrained fault that happens at bus 7. This shows that the safe ARS algorithm can result in emergency control policy adaptive to the fault not seen during the training.}
    \label{fig:safeARS}
\end{figure}

To test the adaptation of the safe ARS, we consider a fault not encountered during the training, in which short-circuit fault happens at bus $7$ at $1.0$s with a fault duration of $0.15$s and then the fault self-cleared. The performances of standard ARS-based load shedding and safe ARS-based load shedding are depicted in Figs. \ref{fig:ARS} and \ref{fig:safeARS}. From these figures, we can observe that the safe ARS-based load shedding is better than the standard ARS-based load shedding not only  in meeting the safety requirement described by the transient voltage recovery criterion, but also in adapting to a fault not encountered during the training.

\section{Conclusions}

In this paper, we presented a highly scalable and safe deep reinforcement learning algorithm for power system voltage stability control using load shedding. This algorithm inherits the parallelism of the ARS algorithm and thereby achieve high scalability and applicability for power system stability control applications. Furthermore, by incorporating the safety function into the reward function, the safe ARS algorithm resulted in a control policy that could enhance the safety of the electric power grid during the load shedding. A small number of hyper-parameters makes this algorithm easy to tune to achieve good performance. Case studies on the IEEE $39$-bus demonstrated that safe ARS-based load shedding successfully recovers the voltage of power systems even in events it did not see during the training. In addition, in comparison to the standard ARS-based load shedding, it showed advantages in both safety level and fault adaptability.

\bibliographystyle{IEEEtran}
\bibliography{ref.bib}
\end{document}